\documentclass[aps,pra, twocolumn, amsmath, amssymb, showpacs, superscriptaddress]{revtex4}

\usepackage{graphicx}% Include figure files
\usepackage{bm}% bold math
\begin{document}

\title{Dynamic steady-state of periodically-driven quantum systems}

\author{V. I. Yudin}
\email{viyudin@mail.ru}
\affiliation{Novosibirsk State University, ul. Pirogova 2, Novosibirsk, 630090, Russia}
\affiliation{Institute of Laser Physics SB RAS, pr. Akademika Lavrent'eva 13/3, Novosibirsk, 630090, Russia}
\affiliation{Novosibirsk State Technical University, pr. Karla Marksa 20, Novosibirsk, 630073, Russia}
%\affiliation{Russian Quantum Center, Skolkovo, Moscow Reg., 143025, Russia}
\author{A. V. Taichenachev}
\affiliation{Novosibirsk State University, ul. Pirogova 2, Novosibirsk, 630090, Russia}
\affiliation{Institute of Laser Physics SB RAS, pr. Akademika
Lavrent'eva 13/3, Novosibirsk, 630090, Russia}
%\affiliation{Russian Quantum Center,
%Skolkovo, Moscow Reg., 143025, Russia}
%
\author{M. Yu. Basalaev}
\affiliation{Novosibirsk State University, ul. Pirogova 2, Novosibirsk, 630090, Russia}
\affiliation{Institute of Laser Physics SB RAS, pr. Akademika Lavrent'eva 13/3, Novosibirsk, 630090, Russia}
\affiliation{Novosibirsk State Technical University, pr. Karla Marksa 20, Novosibirsk, 630073, Russia}
\author{D. Kovalenko}
\affiliation{Novosibirsk State Technical University, pr. Karla Marksa 20, Novosibirsk, 630073, Russia}

%\date{\today}

\begin{abstract}
Using the density matrix formalism, we prove an
existence theorem of the periodic steady-state for an arbitrary
periodically-driven system. This state has the same period as the
modulated external influence, and it is realized as an asymptotic
solution ($t$$\to$$+\infty$) due to relaxation
processes. The presented derivation simultaneously contains a
simple computational algorithm non-using both Floquet and  Fourier
theories, i.e. our method automatically guarantees a full account
of all frequency components. The description is accompanied by the examples demonstrating a simplicity and high efficiency of our method. In particular, for three-level $\Lambda$-system we calculate the lineshape and field-induced shift of the dark resonance formed by the field with periodically modulated phase. For two-level atom we obtain the analytical expressions for signal of the direct frequency
comb spectroscopy with rectangular light pulses. In this case it
was shown the radical dependence of the spectroscopy lineshape on
pulse area. Moreover, the existence of
quasi-forbidden spectroscopic zones, in which the Ramsey fringes
are significantly reduced, is found. The obtained results have a wide area
of applications in the laser physics and spectroscopy, and they
can stimulate the search of new excitation schemes for atomic
clock. Also our results can be useful for many-body physics.
\end{abstract}


\maketitle

For the last several decades a rapid
scientific-and-technological progress was substantially connected
with an expansion of the lasers and laser technologies at
different areas of the science, engineering and industry. In this
process the laser physics, nonlinear optics
and laser spectroscopy take a special place. Many impressive successes in these
directions are due to the theoretical support, motivation and
interpretation of experimental researches. In this context a
paramount importance has the formulation of mathematical models
(equations) and finding of their solutions, which adequately
describe the physical picture of investigated problems. As an
example, for atomic mediums the density matrix formalism is the most
widespread approach describing the atom-field interaction and
different relaxation processes (spontaneous, collisional, etc.).
Especial significance has so-called steady-state, which arises
under the interaction of quantum system with stationary external
fields. During long time this state plays a key role for
theoretical description of the basic
problems in the laser physics and spectroscopy
\cite{Demtroder,Letokhov,Rautian}.

However, in the last years the devices, in which different
parameters of electromagnetic fields are periodically modulated, have gotten a great
importance.
First of all, it is so-called frequency comb-generators using the
periodic pulse modulation of the laser field. Such sources of the
pulse radiation are actively used now in the modern atomic clocks
for frequency measurements \cite{hall06,hansch06}, and also they
have promising perspectives for the direct frequency comb
spectroscopy \cite{vitanov01,stowe08,adler10}. Also it is widely used now
the phase (frequency) and/or amplitude periodic modulation of the
laser field to form the reference resonances for small-size atomic
clocks and magnetometers
\cite{knappe05,post05,phillips05,gawlik06,acosta06,schwindt07}.
Other promising and interesting direction in atomic clocks and
magnetometers is connected with periodically modulated
polarization of the laser field
\cite{jau04,jau07,pustelny06,ben-kish10,huang12,breschi14}. Of
late years the understanding of periodically-driven systems is
also one of the most active areas of research in many-body physics
(e.g., see
\cite{russomanno13,russomanno13_2,D'Alessio13,lazarides14,citro15}).
In all above examples the {\em standard} conception of
steady-state is inapplicable, generally speaking, due to a regular
time variation of external influence.

In this paper, we generalize the steady-state conception for an
arbitrary quantum system under any periodic external influence. In
this way we prove the following existence theorem: if the
coefficients of density matrix dynamic equations have the period
$T$, then the periodic solution with the same period $T$ exists
always. Due to the relaxation processes this solution is realized
as an asymptotics ($t$$\to$$+\infty$) and therefore can be
characterized as periodic steady-state. It is found the simple
algorithm, which allows us directly to construct this solution
independently of initial conditions and without the use of both Floquet and
Fourier theories. Our approach considerably simplifies the
analysis of the atom-field interaction regardless of periodic
modulation character: from smoothly harmonic type to the
ultrashort pulses. As a bright example, for two-level atoms we
analytically calculate the signal of the direct frequency comb
spectroscopy formed by the periodic sequence of rectangular
pulses. The obtained results have great perspectives in the laser
physics and spectroscopy, and they can stimulate the search of
new excitation schemes for atomic clock.

The general mathematical formulation of the problem is following. For an arbitrary quantum system (e.g., atoms) let us consider the density matrix $\hat{\rho}(t)$ and its normalization condition, which in the basis of states $\{ |j\rangle\}$ can be presented as:
\begin{equation}\label{rho}
\hat{\rho}(t)=\sum_{a,b}|a\rangle \rho_{ab}^{}(t)\langle b|;\quad \text{Tr}\{ \hat{\rho}(t)\}=\sum_{j}\rho^{}_{jj}(t)=1 ,
\end{equation}
where $\rho_{ab}^{}(t)$ are the matrix elements. The dynamic equation for density matrix we reformulate to the vector form:
\begin{equation}\label{initial_vec}
\partial_t\vec{\rho}(t)=\widehat{L}(t)\,\vec{\rho}(t)\,; \quad \text{Tr}\{ \hat{\rho}(t)\}=\left(\vec{n},\vec{\rho}(t)\right)=1\, ,
\end{equation}
where the linear operator $\widehat{L}(t)$ describes the interaction with external fields and different relaxation processes (spontaneous, collisional, etc.), the column-vector $\vec{\rho}(t)$ is formed by the matrix elements $\rho_{ab}^{}(t)$ using some definite rule. As an example, for two-level system with basis states $|1\rangle$ and $|2\rangle$ there are four matrix elements $\rho_{ab}^{}(t)$ ($a,b$=1,2), which can be ordered as following vector:
\begin{equation}\label{rho_vec}
\vec{\rho}(t)=
\left(
  \begin{array}{c}
    \rho_{11}^{}(t) \\
    \rho_{12}^{}(t) \\
    \rho_{21}^{}(t) \\
    \rho_{22}^{}(t) \\
  \end{array}
\right) .
\end{equation}
Besides we define the supplementary column-vector $\vec{n}$
permitting to express the value $\text{Tr}\{ \hat{\rho}(t)\}$ as
the dot product in Eq.~(\ref{initial_vec}), where we use the
standard definition of the dot product for arbitrary complex-valued
vectors: $(\vec{x},\vec{y})=\sum_{m}x_{m}^{*}y_{m}^{}$. The vector
$\vec{n}$ is formed in the following way: for the positions
corresponding to the diagonal elements $\rho_{jj}^{}(t)$ [as
components of the vector $\vec{\rho}(t)$] the vector $\vec{n}$ has
the value 1, and it equals to 0 for all other positions. In
particular, for two-level system according to the definition
(\ref{rho_vec}) we obtain:
\begin{equation}\label{n_vec}
\vec{n}=
\left(
  \begin{array}{c}
    1 \\
    0 \\
    0 \\
    1 \\
  \end{array}
\right) .
\end{equation}
The normalization condition in (\ref{initial_vec}) implies the degeneracy of equations [right-hand members, $\widehat{L}(t)\vec{\rho}(t)$] that corresponds to the following expression:
\begin{equation}\label{norm}
\left(\vec{n},\widehat{L}(t)\vec{\rho}\right) =0
\end{equation}
for an arbitrary vector argument $\vec{\rho}$. Note that the
existence of the time-independent vector $\vec{n}$ is a base point
for the next reasonings.

Let us suppose an existence of the time period $T$ in the operator $\widehat{L}(t)$:
\begin{equation}\label{period_L}
\widehat{L}(t+T)=\widehat{L}(t)\, .
\end{equation}
In this case, as it will be shown below, the equation (\ref{initial_vec}) always has a periodic solution with the same period $T$:
\begin{equation}\label{period_rho}
\vec{\rho}(t+T)=\vec{\rho}(t)\,.
\end{equation}

To begin the prove of this theorem, we assume that at some instant of time
$t_{1}^{}$ we have an arbitrary vector $\vec{\rho}(t_{1}^{})$.
Then, in accordance with Eq.~(\ref{initial_vec}), for other
instant of time $t_{2}^{}$ we can write:
\begin{equation}\label{t2t1}
\vec{\rho}(t_{2}^{})=\widehat{A}(t_{2}^{},t_{1}^{})\,\vec{\rho}(t_{1}^{})\,,
\end{equation}
where the two-time evolution operator
$\widehat{A}(t_{2}^{},t_{1}^{})$ is determined by the matrix
$\widehat{L}(t)$. Note that in the case of periodicity condition
(\ref{period_L}) we have the following relationship
\begin{equation}\label{t2Tt1T}
\widehat{A}(t_{2}^{}+ T,t_{1}^{}+ T)=\widehat{A}(t_{2}^{},t_{1}^{})
\end{equation}
for arbitrary $t^{}_{1},t^{}_{2}$.

However, before the consideration of periodic case, let us prove
several general statements. For this purpose, we multiply the
Eq.~(\ref{t2t1}) by the vector $\vec{n}$ as the dot product:
\begin{equation}\label{nrt2t1}
\left(\vec{n},\vec{\rho}(t_{2}^{})\right)=\left(\vec{n},\widehat{A}(t_{2}^{},t_{1}^{})\,\vec{\rho}(t_{1}^{})\right)=
\left(\widehat{A}^{\dagger}(t_{2}^{},t_{1}^{})\,\vec{n},\vec{\rho}(t_{1}^{})\right),
\end{equation}
where $\widehat{A}^{\dagger}(t_{2}^{},t_{1}^{})$ is the Hermitian
conjugate operator to the $\widehat{A}(t_{2}^{},t_{1}^{})$.
Because of
$(\vec{n},\vec{\rho}(t_{2}^{}))$=$(\vec{n},\vec{\rho}(t_{1}^{}))$,
from Eq.~(\ref{nrt2t1}) we obtain:
\begin{equation}\label{nrt1}
\left(\vec{n},\vec{\rho}(t_{1}^{})\right)=\left(\widehat{A}^{\dagger}(t_{2}^{},t_{1}^{})\,\vec{n},\vec{\rho}(t_{1}^{})\right)
\end{equation}
for arbitrary $\vec{\rho}(t_{1}^{})$. This formula directly implies
\begin{equation}\label{An}
\widehat{A}^{\dagger}(t_{2}^{},t_{1}^{})\,\vec{n}=\vec{n}\,,
\end{equation}
i.e. the operator $\widehat{A}^{\dagger}(t_{2}^{},t_{1}^{})$ always has the eigenvector with the real eigenvalue 1. Therefore, the same eigenvalue 1 also exists for the direct operator $\widehat{A}(t_{2}^{},t_{1}^{})$, i.e. always there is the eigenvector $\vec{r}(t_{2}^{},t_{1}^{})$, which satisfies the equation:
\begin{equation}\label{Ar}
\widehat{A}(t_{2}^{},t_{1}^{})\,\vec{r}(t_{2}^{},t_{1}^{})=\vec{r}(t_{2}^{},t_{1}^{})\,.
\end{equation}
In contrast to Eq.~(\ref{An}), the eigenvector
$\vec{r}(t_{2}^{},t_{1}^{})$ depends on $t_{2}^{},t_{1}^{}$ in the
general case.

Let us return to the case of periodicity (\ref{period_L}).
Consider the  $\vec{\rho}(t)$ at arbitrary instant of time $t$. In
conformity with Eq.~(\ref{t2t1}), the vector $\vec{\rho}(t+T)$ is determined as
\begin{equation}\label{tT}
\vec{\rho}(t+T)=\widehat{A}(t+T,t)\,\vec{\rho}(t)\,.
\end{equation}
Supposing the existence of the periodic solution $\vec{\rho}(t+T)$=$\vec{\rho}(t)$, it follows from Eq.~(\ref{tT}) that this solution satisfies the equation:
\begin{equation}\label{tT_period}
\vec{\rho}(t)=\widehat{A}(t+T,t)\,\vec{\rho}(t)\,;\quad (\vec{n},\vec{\rho}(t))=1\,,
\end{equation}
which always has nonzero solution due to the above proven statement of existence of eigenvector with eigenvalue 1 for the operator $\widehat{A}(t_{2}^{},t_{1}^{})$ for arbitrary $t^{}_{1},t^{}_{2}$ [see Eq.~(\ref{Ar})]. Using the Eqs.~(\ref{t2Tt1T}) and (\ref{tT}),(\ref{tT_period}), it can be easily shown that $\vec{\rho}(t+lT)$=$\vec{\rho}(t)$ ($l$=$\pm 1,\pm 2, ...$). Taking into account an arbitrariness of the time $t$ in the Eqs.~(\ref{tT}),(\ref{tT_period}), we can assert that the existence theorem of the periodic solution (\ref{period_rho}) is proven.

Due to the relaxation processes there is an unique (in the majority of cases) periodic solution, which is realized during the time evolution as an asymptotic state ($t$$\to$$+\infty$) independently of initial conditions (similarly to the well-known case of stationary field (\ref{steady_state}) -- see below). The solution (\ref{period_rho}) can be called as {\em dynamic} steady-state because of its dependence on $t$.

However, taking into account a mathematical generality, one can theoretically assume the possibility of several solutions \{$\vec{\rho}^{}_{1}(t)$,$\vec{\rho}^{}_{2}(t)$,...,$\,\vec{\rho}^{}_{Q}(t)$\} for Eq.~(\ref{tT_period}), when the eigenvalue 1 for the matrix $\widehat{A}(t+T,t)$ is degenerated. In this case the general periodic solution has a form of superposition:
\begin{equation}\label{superposition}
\vec{\rho}(t)=\sum_{q =1}^{Q}\alpha_{q}\vec{\rho}_{q}(t)\,;\quad (\vec{n},\vec{\rho}_{q}(t))=1\,;\quad \sum_{q =1}^{Q}\alpha_{q}=1\,,
\end{equation}
where the number set \{$\alpha^{}_{1}$,...,$\,\alpha^{}_{Q}$\} for the asymptotic solution ($t\to +\infty$) will depend on initial conditions (for the dynamical consideration).

\begin{figure}[t]
\centerline{\scalebox{0.45}{\includegraphics{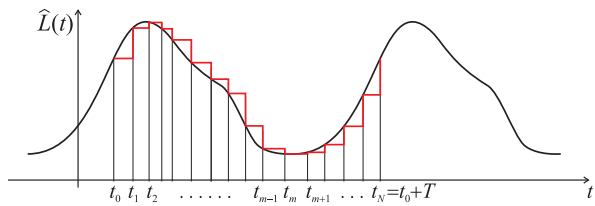}}}\caption{
The partition of the time interval $[t_0,t_0+T]$ at $N$
subintervals and symbolic approximation of the dependence
$\widehat{L}(t)$ by the step function (red step line).} \label{Fig1}
\end{figure}

Let us show that the found solution (\ref{tT_period}) is a
generalization of well-known steady-state for the constant
operator $\widehat{L}$ in Eq.~(\ref{initial_vec}):
\begin{equation}\label{steady_state}
\widehat{L}\vec{\rho}_\text{st-st}=0\,;\quad
\partial_t\vec{\rho}_\text{st-st}=0\,; \quad
(\vec{n},\vec{\rho}_\text{st-st})=1.
\end{equation}
It is obviously that the stationary case can be considered as
periodic case with an arbitrary value of the period $T$.
Therefore, according to the above analysis, the state
$\vec{\rho}_\text{st-st}$ also must satisfy the Eq.~(\ref{tT_period})
for arbitrary $t,T$. Indeed, in the case of $\widehat{L}$=$const$
we have (for arbitrary $t,T$):
\begin{equation}\label{steady_state}
\widehat{A}(t+T,t)=e^{\widehat{L}T}=1+\sum_{k=1}^{+\infty}\frac{1}{k!}T^k\widehat{L}^k\,.
\end{equation}
Taking into account Eq.~(\ref{steady_state}), we obtain the perfect compliance with Eq.~(\ref{tT_period}): $\widehat{A}(t+T,t)\vec{\rho}_\text{st-st}$=$\vec{\rho}_\text{st-st}$.

Let us compare our results with the general Floquet theory (e.g., see \cite{coddington1955}). On this theory the solutions of the differential equation system (\ref{initial_vec}) with periodic coefficients can be presented as following form:
\begin{equation}\label{Floke}
\vec{\rho}(t)=\widehat{D}(t)e^{\widehat{K}t}\vec{a}\,,
\end{equation}
where the matrix $\widehat{D}(t)$=$\widehat{D}(t+T)$ is periodic, $\widehat{K}$ is a certain constant matrix connected with initial matrix $\widehat{L}(t)$, $\vec{a}$ is an arbitrary constant vector. To have the periodic solution the matrix $\widehat{K}$ must satisfy certain criteria. From this viewpoint, it is most probably that in the present paper we have proved the existence of the eigenvector $\vec{a}_0$ with eigenvalue 0 for the matrix $\widehat{K}$: $\widehat{K}\vec{a}_0$=0. In this case, takin into account $e^{\widehat{K}t}\vec{a}_0$=$\vec{a}_0$, from Eq.~(\ref{Floke}) we find the following solution:
\begin{equation}\label{Floke0}
\vec{\rho}(t)=\widehat{D}(t)e^{\widehat{K}t}\vec{a}_0=\widehat{D}(t)\vec{a}_0\,,
\end{equation}
which obviously is periodic [due to the periodicity of $\widehat{D}(t)$]. We believe that exactly this scenario is realized for density matrix equation. However, this problem requires an additional mathematical investigation.

Note that some authors supposed (without proof) the existence of
the periodic solution (\ref{period_rho}) for some certain
problems. In this case they usually used the Fourier analysis for
numerical calculations (e.g., see \cite{post05,moreno14}).
However, as far as we know, the general mathematical proof of the
existence theorem was not previously presented in scientific
literature. At the same time a special interest has the direct and
simple method, which allows us to calculate the periodic solution on
the base of Eq.~(\ref{tT_period}) without Fourier expansion, i.e. our
approach automatically takes account of all frequency components. It
can radically simplify the calculations for
periodically-driven systems.

Let us describe one of possible numerical algorithms based on
Eq.~(\ref{tT_period}). We consider an arbitrary periodic
dependence of the operator $\widehat{L}(t)$ (see Fig.\ref{Fig1}).
For instance, under atom-field interaction such a dependence can
be produced by the modulation of the field parameters (amplitude,
phase, polarization, etc.). The selected time interval
$[t_0,t_0+T]$ we divide into $N$ small subintervals, where
$t_N$=$t_0+T$. The character of partition (uniform or nonuniform
discrete mesh) and number of subintervals are determined in
conformity with studied problem. The dependence $\widehat{L}(t)$
we will approximate by step function (see Fig.~\ref{Fig1}), where
the matrix $\widehat{L}(t)$ has the constant value
$\widehat{L}(t_{m-1})$ inside of subinterval $(t_{m-1},t_{m}]$. In
this case the vector $\vec{\rho}(t_0)$ in initial point $t_0$ is
determined by Eq.~(\ref{tT_period}), where the evolution operator
$\widehat{A}(t_0+T,t_0)$ has the form of an ordered product of the
matrix exponents:
\begin{align}\label{Matrix_Exp0}
\widehat{A}(t_0+T,t_0)=&\prod_{m=1}^{m=N}e^{(t_{m}-t_{m-1})\widehat{L}(t_{m-1})}=\\
& e^{(t_{N}-t_{N-1})\widehat{L}(t_{N-1})}\ldots
e^{(t_{1}-t_{0})\widehat{L}(t_{0})}.\nonumber
\end{align}
The vectors $\vec{\rho}(t_m)$ in other points of the interval $[t_0,t_0+T]$ are determined by the recurrence relation:
\begin{equation}\label{Matrix_Exp_m}
\vec{\rho}(t_{m})=e^{(t_{m}-t_{m-1})\widehat{L}(t_{m-1})}\vec{\rho}(t_{m-1})\,.
\end{equation}

\begin{figure}[t]
\centerline{\scalebox{0.38}{\includegraphics{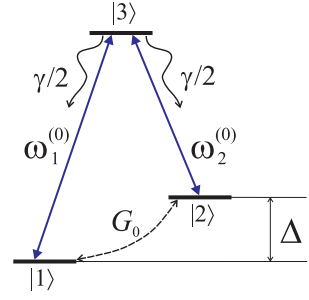}}}\caption{
Three-level system where the dark resonance can be formed.} \label{scheme_CPT}
\end{figure}

Below we give two examples, which demonstrate the simplicity and high efficiency of our method. First example is the so-called dark resonance, which is formed in three-level $\Lambda$-system by bichromatic field (see Fig.~\ref{scheme_CPT}). This resonance occurs when a difference between the optical frequencies ($\omega_1-\omega_2$) is varied near the transition frequency between the lower energy levels $|1\rangle$ and $|2\rangle$: $\omega_1-\omega_2\approx\Delta$. Currently, such resonances at the hyperfine structure of the alkali metal atoms (first of all, for $^{87}$Rb and $^{133}$Cs) underlie the chip-scale atomic clocks, which have a great practical importance. At the same time the harmonic phase modulation (frequency deviation) of miniature semiconductor lasers (so-called VICSEL) is ordinarily used in practice. In this case the field can be written as
\begin{equation}\label{Phase_mod}
E(t)=E_0 e^{-i[\omega t+\varphi(t)]}+c.c.\,;\quad \varphi(t)=A\sin(\nu t)\,,
\end{equation}
where $E_0$ is the field amplitude, $\omega$ is the central frequency of laser, $A$ and $\nu$ are the amplitude and frequency of phase modulation, respectively. It is obvious that this field is polychromatic, in which the frequency difference between the adjacent components equals to $\nu$. The phase modulation at full- or half of the hyperfine splitting is usually used: $\nu$$\approx$$\Delta$ or $\Delta/2$. However, the dark resonance can be observed in the more general case of $\nu$$\approx$$\Delta/l$ ($l$=1,2,3,...).

For theoretical description of this problem the Fourier expansion of the field (\ref{Phase_mod}) is traditionally used. In this case, to solve the density matrix equation one can extract two resonant components (e.g., see \cite{post05}), which are involved in the absorption and form the dark resonance per se. While the remaining frequency components are taken into account only from the viewpoint of field-induced shifts of the clock transition. It is evident, that such approach is quite cumbersome and incomplete. In our method the Fourier expansion is not used and we directly calculate the time-dependent periodic $\hat{\rho}(t)$. In the resonance approximation the equations for the density matrix components $\rho^{}_{jk}(t)$ have the following form:
\begin{align}\label{rho_Lambda}
&[\partial_t+\Gamma-i\delta_1-i\varphi ' (t)]\rho_{31}=i\Omega^{}_1(\rho_{11}-\rho_{33})+i\Omega^{}_2\rho_{21}\nonumber \\
&[\partial_t+\Gamma-i\delta_2-i\varphi ' (t)]\rho_{32}=i\Omega^{}_2(\rho_{22}-\rho_{33})+i\Omega^{}_1\rho_{12}\nonumber\\
&[\partial_t+G_0 -i\Delta]\rho_{12}=i(\Omega^{\ast}_1\rho_{32}-\rho_{13}\Omega^{}_{2})\\
&[\partial_t+G_0 ] \rho_{11}=\gamma\rho_{33}/2+G_0 \text{Tr}\{\hat{\rho}\}/2+i(\Omega^{\ast}_1\rho_{31}-\rho_{13}\Omega^{}_{1})\nonumber\\
&[\partial_t+G_0 ] \rho_{22}=\gamma\rho_{33}/2+G_0 \text{Tr}\{\hat{\rho}\}/2+i(\Omega^{\ast}_2\rho_{32}-\rho_{23}\Omega^{}_{2})\nonumber\\
&[\partial_t+G_0 +\gamma] \rho_{33}= i(\Omega^{}_1\rho_{13}-\rho_{31}\Omega^{\ast}_{1})+i(\Omega^{}_2\rho_{23}-\rho_{32}\Omega^{\ast}_{2})\nonumber\\
&\rho^{}_{jk}=\rho^{\ast}_{kj}\;(j,k=1,2,3);\;\; \text{Tr}\{\hat{\rho}\}=\rho_{11}+\rho_{22}+\rho_{33}=1.\nonumber
\end{align}
Here $\delta_{1,2}$=($\omega-\omega^{(0)}_{1,2}$) are one-photon detunings,  $\Omega_1$=$d_{31}E_0$/$\hbar$ and $\Omega_2$=$d_{32}E_0$/$\hbar$ are the Rabi frequencies ($d_{31}$ and $d_{32}$ are the reduced matrix elements of the dipole moment for the respective transitions), $\gamma$ is decay rate of excited level $|3\rangle$, $\Gamma$ is the total decoherence rate (spontaneous, collision, time-of-flight) of optical transitions $|1\rangle$$\to$$|3\rangle$ and $|2\rangle$$\to$$|3\rangle$, $G_0 $ is relaxation rate of lower energy levels to the equilibrium  isotropic state: $(|1\rangle\langle 1|+|2\rangle\langle 2|)/2$. Note, if we delete the derivative by phase modulation $\varphi '(t)$, then the equations (\ref{rho_Lambda}) formally coincide with the description of interaction between monochromatic field (at the frequency $\omega$) and three-level $\Lambda$-system.

\begin{figure}[t]
\centerline{\scalebox{0.37}{\includegraphics{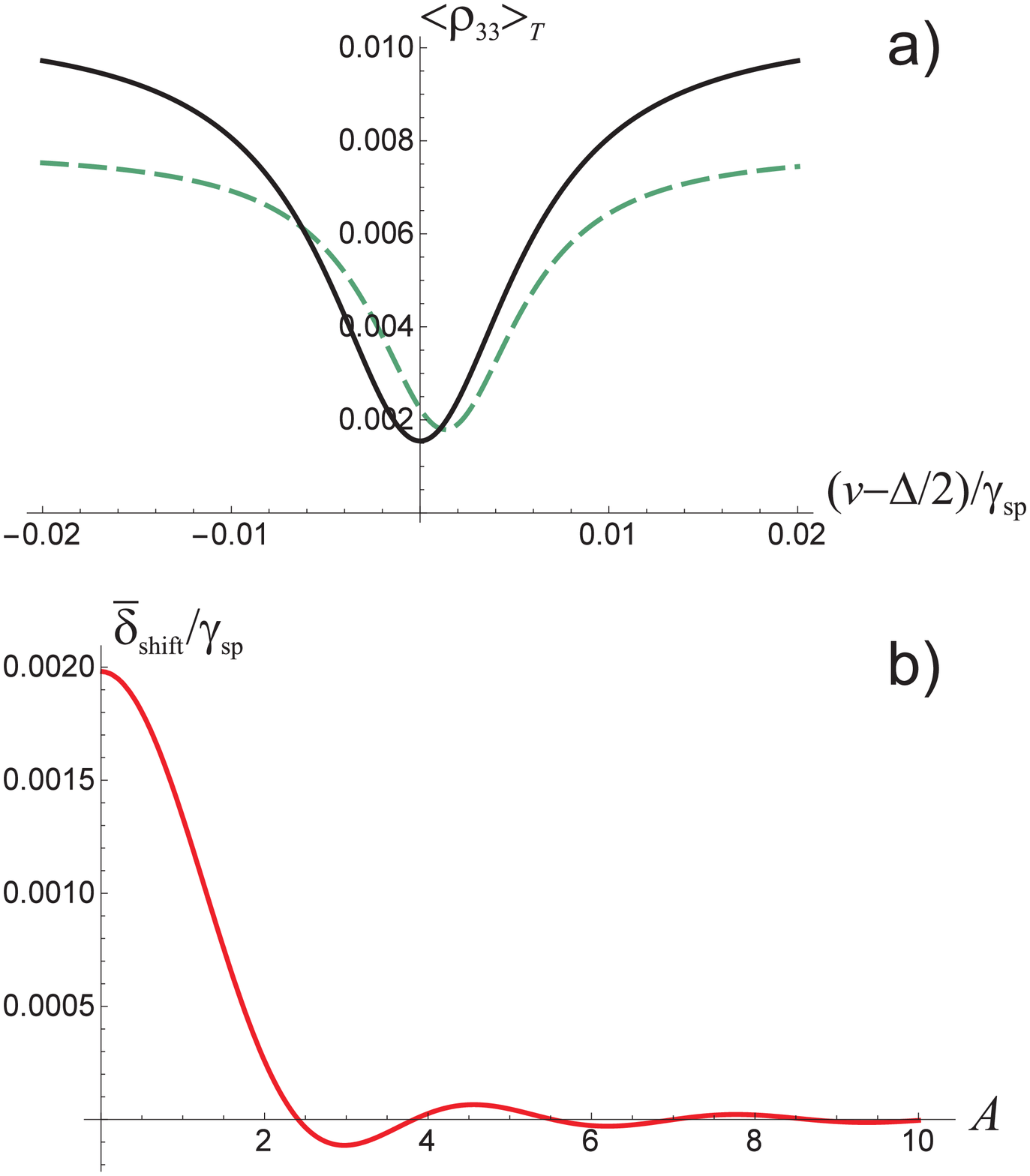}}}\caption{
a) The dark resonance lineshape in the case when $\nu$ is varied near $\Delta/2$ for different values of the phase modulation amplitude $A$: $A$=1.0 (green dashed line) and $A$=2.4 (solid black line); b) The dependance of the dark resonance position (relative to $\Delta/2$) on parameter $A$. \\
Calculations are done for $\Omega_1/\gamma $=$\Omega_2/\gamma $=1; $G_0 /\gamma $=1.5$\times$$10^{-3}$; $\Gamma/\gamma $=50; $\Delta/\gamma $=1000, $\delta_2$=$-\delta_1$=$\Delta/2$.}
\label{rho33}
\end{figure}

\begin{figure}[h]
\centerline{\scalebox{0.37}{\includegraphics{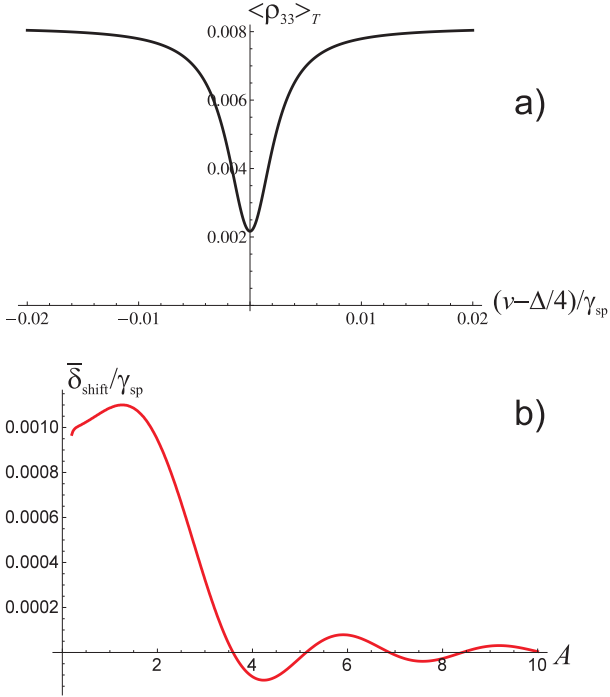}}}\caption{
a) The dark resonance lineshape in the case when $\nu$ is varied near $\Delta/4$ for the phase modulation amplitude $A$=3.59; b) The dependance of the dark resonance position (relative to $\Delta/4$) on parameter $A$. \\
Calculations are done for $\Omega_1/\gamma $=$\Omega_2/\gamma $=1; $G_0 /\gamma $=1.5$\times$$10^{-3}$; $\Gamma/\gamma $=50; $\Delta/\gamma $=1000, $\delta_2$=$-\delta_1$=$\Delta/2$.}
\label{rho33_2}
\end{figure}

In particular, we have calculated by our method the resonance lineshapes as dependencies on the frequency of phase modulation $\nu$$\approx$$\Delta/2$ for different values of amplitude $A$ (see Fig.~\ref{rho33}a). The taken parameters are typical for atomic cells with vapor of  $^{87}$Rb in buffer gas. Fig.~\ref{rho33}b displays the dependence of field shift of dark resonance $\bar{\delta}_\text{shift}$ (with respect to $\Delta$/2) on modulation amplitude $A$. There is the value set of $A$, for which the field shift vanishes (see also \cite{zhu2000}). In Fig.~\ref{rho33_2} the similar graphics are shown for the case of phase modulation frequency $\nu$$\approx$$\Delta/4$. From Fig.~\ref{rho33_2}a it is seen that the amplitude of dark resonance can be comparable with the case of $\nu$$\approx$$\Delta/2$ (compare with Fig.~\ref{rho33}a).

\begin{figure}[t]
\centerline{\scalebox{0.35}{\includegraphics{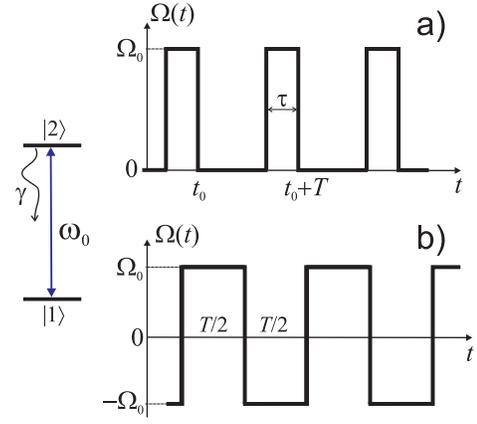}}}\caption{
Two-level system interacting with: a) periodic sequence of
rectangular pulses; b) periodic sequence of $\pm\pi$ phase jumps.} \label{schemes}
\end{figure}

To show other examples, let us consider two-level system (with unperturbed
frequency $\omega^{}_{0}$) interacting with the field at the
frequency $\omega$: $E(t)$=$\text{Re}\{ {\cal E}(t)\,e^{-i\omega
t}\}$, where the amplitude modulation ${\cal E}(t)$ has the form
of rectangular-pulse periodic sequence (see Fig.~\ref{schemes}a), which can be considered as limiting case of the Ramsey spectroscopy. In this case the field has an equidistant spectrum
$\omega_m$=$\omega+mf_r$ ($m$=0,$\pm 1$,$\pm 2$...) (so-called
frequency comb), where $f_r$=$2\pi /T$ is the repetition
frequency, and spectrum width (with respect to the central
frequency $\omega$) is determined by the pulse duration as
$1/\tau$. The dynamic equations for density matrix components are
given by:
\begin{align}\label{2_level}
&[\partial_t+\Gamma-i\delta]\rho_{21}=i\Omega(t)(\rho_{11}-\rho_{22})/2\nonumber \\
&[\partial_t+\Gamma+i\delta]\rho_{12}=-i\Omega^{\ast}(t)(\rho_{11}-\rho_{22})/2\nonumber \\
&[\partial_t+\gamma ] \rho_{22}=i[\Omega(t)\rho_{12}-\rho_{21}\Omega^{\ast}(t)]/2\\
&\partial_t\rho_{11}=\gamma \rho_{22}+i[\Omega^{\ast}(t)\rho_{21}-\rho_{12}\Omega^{}(t)]/2\nonumber\\
&\text{Tr}\{\hat{\rho}\}=\rho_{11}+\rho_{22}=1\,,\nonumber
\end{align}
where $\Omega(t)$=$d{\cal E}(t)$/$\hbar$ is the time-dependent Rabi frequency ($d$ is the dipole matrix element), $\delta$=($\omega-\omega^{}_{0}$) is the detuning of the central frequency $\omega$. The constants $\gamma$ and $\Gamma$ describe the relaxation rates of the exited state population ($\rho_{22}$) and optical coherence ($\rho_{12},\rho_{12})$, respectively. Note that for pure spontaneous relaxation the condition
$\Gamma$=$\gamma /2$ takes place.

\begin{figure}[h]
\centerline{\scalebox{0.55}{\includegraphics{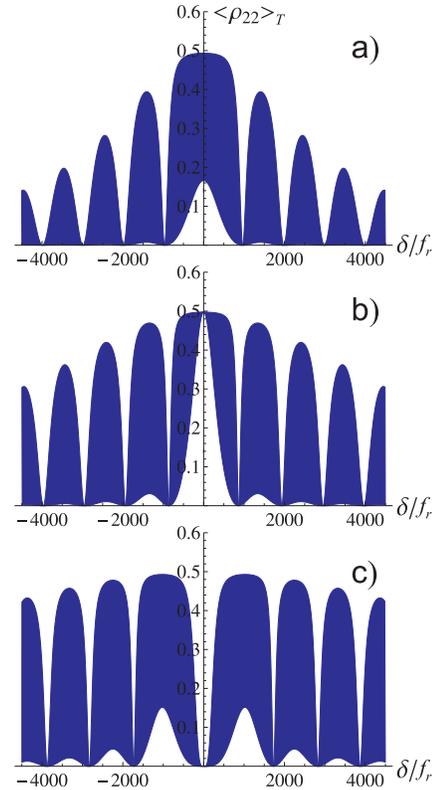}}}\caption{
The lineshape for direct frequency comb spectroscopy (see
Fig.~\ref{schemes}a) over large interval of detuning $\delta$,
where the narrow resonances are visually inseparable (dark
areas). Calculations are done for $\Gamma$=$\gamma
/2$=0.02$f_r$, $\tau /T$=0.001 and for different pulse area:
a) $\Omega_0\tau$=$\pi$/2; b) $\Omega_0\tau$=$\pi$;
c) $\Omega_0\tau$=2$\pi$.} \label{comb_full}
\end{figure}

\begin{figure}[h]
\centerline{\scalebox{0.7}{\includegraphics{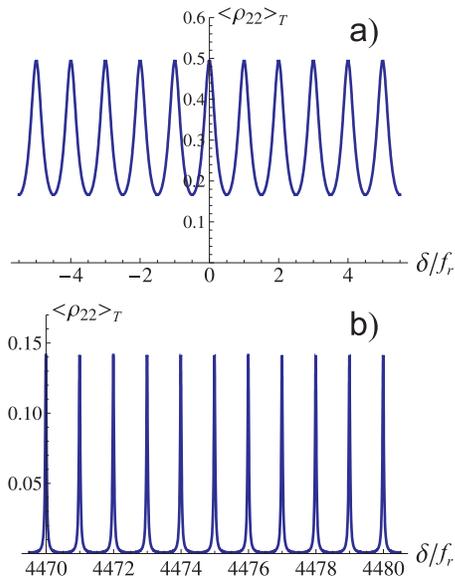}}}\caption{
The lineshapes for direct frequency comb spectroscopy (see
Fig.~\ref{schemes}a) over two small intervals of detuning $\delta$,
where the individual Ramsey resonances are visually separated. Calculations
are done for: $\Gamma$=$\gamma /2$=0.02$f_r$, $\tau
/T$=0.001, $\Omega_0\tau$=$\pi$/2.} \label{comb_part}
\end{figure}

The most interest has a regime, in which the repetition frequency
significantly exceeds the relaxation rate of atomic system:
$\Gamma$,$\gamma $$\ll$$f_r$, and also the realization
of the condition $\Omega_0\tau$$\geq$1. The
Figs.~\ref{comb_full}a-c demonstrate the dependences of
period-averaged value $\langle\rho_{22}\rangle^{}_T$ as a function
of detuning $\delta$ for different pulse area
$\Omega_0\tau$=$\pi/2$,$\,\pi$,$\,2\pi$. The structure of these
lineshapes consist of large number of resonances at the
frequencies $\omega$=$\omega_0$$+mf_r$ ($m$=0,$\pm 1$,$\pm
2$,...). Their amplitudes lie between two envelope curves. The
dark areas in Fig.~\ref{comb_full} correspond to the resonances,
which are visually inseparable for given scale. More detail
picture using small frequency intervals is presented in
Fig.~\ref{comb_part}, where the individual resonances are visible.
We have found that the lower envelope has a typical width of the
order of 2$\Omega_0$ (see in Fig.~\ref{comb_full}) and it looks
similar to the lineshape on $\delta$ in the case of single-pulse
Rabi spectroscopy. The upper envelope is much more broadened  (with width of the order
of $\Omega_0 f_r$$/\gamma $$\gg$$\Omega_0$ in the case of $\Omega_0\tau$$\geq$1).
Moreover, this envelope curve has a series of
relatively narrow resonance-like dips (practically to zero), which
can be characterized as quasi-forbidden spectroscopic zones. The
positions of these zones are determined from the condition:
$\tau\sqrt{\Omega_0^2+\delta^2}$=2$k\pi$ ($k$=1,2,...), and their
widths are proportional to $\gamma$.

Note the high efficiency and simplicity of our method in the case of periodic pulse modulation. Indeed, for above considered task with rectangular pulses (see Fig.~\ref{schemes}a) the operator $\widehat{A}(t_0+T,t_0)$ includes the product of only two matrix exponents:
\begin{equation}\label{MExp2}
\widehat{A}(t_0+T,t_0)=e^{\tau\widehat{L}(\Omega=\Omega_0)}\,e^{(T-\tau)\widehat{L}(\Omega=0)}\,.
\end{equation}
In result, the calculation of the dynamic steady-state in concordance with (\ref{tT_period}) becomes quite simple. Moreover, for two-level atom we have obtained the analytical expressions (exact and approximative) for the operator $\widehat{A}(t_0+T,t_0)$ and vector $\vec{\rho}(t_0)$. In particular, for the short pulses ($\tau$$\ll$$1/\gamma$, $\tau$$\ll$$T$) the following approximation takes place:
\begin{align}\label{analit}
& \rho_{22}(t_0)\approx \Omega_0^2 e^{\gamma {\cal T}}(e^{2\Gamma{\cal T}}-1)\sin^2(\Omega\tau/2)\times \nonumber\\
& \{ (e^{(2\Gamma+\gamma){\cal T}}-1)\Omega^2+ 2e^{\Gamma{\cal T}}(e^{\gamma {\cal T}}-1) \Omega\delta\sin(\delta {\cal T})\sin(\Omega\tau)- \nonumber\\
&  e^{\Gamma{\cal T}}(e^{\gamma {\cal T}}-1)\cos(\delta {\cal T})[\Omega_0^2+(2\delta^2+\Omega_0^2)\cos(\Omega\tau)]+ \nonumber \\
& (e^{\gamma {\cal T}}-e^{2\Gamma {\cal T}})[\Omega_0^2\cos(\Omega\tau)+\delta^2]\}^{-1},
\end{align}
where ${\cal T}$=($T-\tau$) is a free evolution time between
pulses, $\Omega$=$\sqrt{\Omega_0^2+\delta^2}$ is the generalized
Rabi frequency. For the pure spontaneous relaxation ($\Gamma$=$\gamma /2$) the expression (\ref{analit}) can be transformed to the form
\begin{align}\label{analit1}
\rho_{22}(t_0)&\approx {\Omega_0^2\exp(\gamma {\cal T}/2)\sin^2(\Omega\tau/2)}\times\nonumber \\
&\{ 2\Omega^2\cosh(\gamma {\cal T}/2)+2\Omega\delta\sin(\delta {\cal T})\sin(\Omega\tau)-\nonumber\\
&  \cos(\delta {\cal T})[\Omega_0^2+(2\delta^2+\Omega_0^2)\cos(\Omega\tau)]\}^{-1}.
\end{align}
In addition, under condition of $\tau$$\ll$$T$ we
have a close approximation:
\begin{equation}\label{approx}
\langle\rho_{22}\rangle^{}_T\approx\rho_{22}(t_0)[1-e^{-\gamma
T}]/\gamma T
\end{equation}
with an absolute accuracy of small value $(\tau/T)$$\ll$1. Thus, to reproduce with a good
accuracy the lineshape of $\langle\rho_{22}\rangle^{}_T$ we can calculate only the dependence $\rho_{22}(t_0)$ on
$\delta$. Note that the Eq.~(\ref{approx}) can be used for an
arbitrary pulse form (not only rectangular).

In contrast, for frequency comb spectroscopy to numerically solve the Eq.~(\ref{initial_vec}) using the Fourier analysis we should use the following decomposition:  $\vec{\rho}(t)=\sum_{m}\vec{\rho}_m e^{mf_rt}$ ($m=0,\pm 1,\pm 2,...$),
where the components $\vec{\rho}_m$ satisfy the certain recurrent relations. As an example, the calculations of the dependencies in Fig.~\ref{comb_full} require  the using of more than 10$^5$ frequency components (for field and density matrix). It leads to the huge computational burden (comparable with supercomputer).

\begin{figure}[h]
\centerline{\scalebox{0.7}{\includegraphics{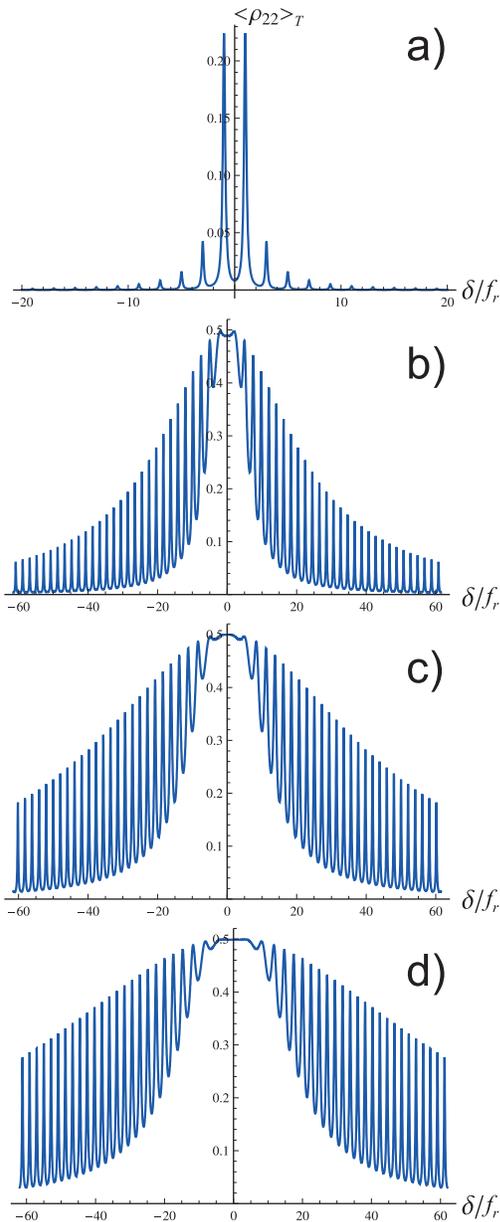}}}\caption{
The spectroscopy lineshapes for the field modulation, which is formed by $\pm\pi$ phase jumps (see Fig.~\ref{schemes}b). Calculations
are done for $\Gamma$=$\gamma /2$=0.1$f_r$ and for different Rabi frequency:
a) $\Omega_0$=0.2$f_r$; b) $\Omega_0$=5$f_r$;
c) $\Omega_0$=10$f_r$; d) $\Omega_0$=15$f_r$.} \label{jumps}
\end{figure}

The last numerical example is shown in Fig.~\ref{jumps}. There are the spectroscopy lineshapes for the field modulation, which is formed by $\pm\pi$ phase jumps (see Fig.~\ref{schemes}b). In the case of $\Omega_0\gg\gamma$, these interesting dependencies have the lower envelope, which has a width of the order of 2$\Omega_0$ (see Fig.~\ref{jumps}b,c,d).

Note that for free atom gas it is necessary to take into account the motion of atoms, which results in dependence of the density matrix on velocity ${\bf v}$. In this case the periodic steady-state $\hat{\rho}(t+T,{\bf v})$=$\hat{\rho}(t,{\bf v})$ should be found for each velocity group, and then the spectroscopic signal is computed with the use of velocity averaging. In addition, the performed in this article consideration of temporal periodic modulation can be adapted for the case of the spatially periodic modulation of the field parameters (amplitude, phase, polarization, etc.). It implies the proof of existence and the calculation of spatially periodic steady-state of the atoms in the spatial lattices formed by electromagnetic field (including the problems of laser cooling and trapping of atoms in optical lattices).

In conclusion, in the frame of density matrix formalism we have
rigorously proven the existence theorem of the periodic
steady-state for an arbitrary periodically-driven system. Due to
the relaxation processes this state is realized as an asymptotics
($t$$\to$$+\infty$) independently of initial conditions. The
presented proof simultaneously contains a computational algorithm,
which uses neither Floquet nor Fourier theories, i.e. our method
automatically takes into account all frequency components. It
radically simplifies the calculations for arbitrary types
of periodic modulation (including the ultrashort pulses) and opens
up great possibilities for analysis and development of new methods
in the laser physics, nonlinear optics and spectroscopy.
The importance of obtained results becomes especially obvious due to
the infinite variety of possible periodic actions for different
quantum systems.

We thank E. Demler for useful discussions. The work was supported by the Ministry of Education and Science of
the Russian Federation (state assignment No. 2014/139, project No.
825), by the Russian Foundation for Basic Research (grants No.
14-02-00712, No. 14-02-00939, No. 15-02-08377, No. 15-32-20330).
M. Yu. Basalaev was also supported by the Dynasty
Foundation.


\begin{thebibliography}{30}
%
\bibitem{Demtroder}
W. Demtroder, {\em Laser Spectroscopy} (Springer-Verlag, Berlin,
2003).
%
\bibitem{Letokhov}
V. S. Letokhov and V. P. Chebotaev, {\em Nonlinear laser
spectroscopy} (Springer-Verlag, Berlin, 1977).
%
\bibitem{Rautian}
S. G. Rautian and A. M. Shalagin, {\em Kinetic problems of
non-linear spectroscopy} (North-Holland, Amsterdam, 1991).
%
\bibitem{hall06}
J. L. Hall, Rev. Mod. Phys. {\bf 78}, 1279 (2006).
%
\bibitem{hansch06}
T. W. H$\ddot{\text{a}}$nsch, Rev. Mod. Phys. {\bf 78}, 1297 (2006).
%
\bibitem{vitanov01}
N. V. Vitanov, M. Fleischhauer, B. W. Shore, K. Bergmann, Adv. At. Mol. Opt. Phys. {\bf 46}, 55 (2001).
%
\bibitem{stowe08}
M. C. Stowe, M. J. Thorpe, A. Pe'er, J. Ye, J. E. Stalnaker, V. Gerginov, and S. A. Diddams, Adv. At. Mol. Opt. Phys. {\bf 55}, 2 (2008).
%
\bibitem{adler10}
F. Adler, M. J. Thorpe, K. C. Cossel, and J. Ye, Annu. Rev. Anal. Chem. {\bf 3}, 175 (2010).
%
\bibitem{knappe05}
S. Knappe, P. D. D. Schwindt, V. Shah, L. Hollberg, J.~Kitching, L. Liew and J. Moreland, Opt. Express {\bf 13}, 1250 (2005).
%
\bibitem{post05}
A. B. Post, Y.-Y. Jau, N. N. Kuzma, and W. Happer, Phys. Rev. A {\bf 72}, 033417 (2005).
%
\bibitem{phillips05}
D. F. Phillips, I. Novikova, Ch. Y.-T. Wang, and R.~L.~Walsworth, J. Opt. Soc. Am. B {\bf 22}, 305 (2005).
%
\bibitem{gawlik06}
N. W. Gawlik, L. Krzemien, S. Pustelny, D. Sangla, J. Zachorowski, M. Graf, A.O. Sushkov, and D. Budker, Appl. Phys. Lett. {\bf 88}, 131108 (2006).
%
\bibitem{acosta06}
V. Acosta, M. P. Ledbetter, S. M. Rochester, D. Budker, D. F. Jackson-Kimball, D. C. Hovde, W. Gawlik, S.~Pustelny, and J. Zachorowski, Phys. Rev. A {\bf 73}, 053404 (2006).
%
\bibitem{schwindt07}
P. D. D. Schwindt, B. Lindseth, S. Knappe, V. Shah, J.~Kitching, and L.-A. Liew, Appl. Phys. Lett. {\bf 90}, 081102 (2007).
%
\bibitem{jau04}
Y.-Y. Jau, E. Miron, A. B. Post, N. N. Kuzma, and W.~Happer, Phys. Rev. Lett. {\bf 93}, 160802 (2004).
%
\bibitem{jau07}
Y.-Y. Jau and W. Happer, Phys. Rev. Lett. {\bf 99}, 223001 (2007).
%
\bibitem{pustelny06}
S. Pustelny, W. Gawlik, S. M. Rochester, D. F. Jackson Kimball,
V. V. Yashchuk, and D. Budker, Phys. Rev. A {\bf 74}, 063420 (2006).
%
\bibitem{ben-kish10}
A. Ben-Kish, and M.V. Romalis, Phys. Rev. Lett. {\bf 105}, 193601 (2010).
%
\bibitem{huang12}
M. Huang, and J. C. Camparo, Phys. Rev. A {\bf 85}, 012509 (2012).
%
\bibitem{breschi14}
E. Breschi, Z. D. Gruji$\acute{\text{c}}$, P. Knowles, and A. Weis, Appl. Phys. Lett. {\bf 104}, 023501 (2014).
%
\bibitem{russomanno13}
A. Russomanno, A. Silva, and G. E. Santoro, Phys. Rev. Lett. {\bf 109}, 257201 (2013).
%
\bibitem{D'Alessio13}
L. D'Alessio, A. Polkovnikov, Annals of Physics {\bf 333}, 19 (2013).
%
\bibitem{russomanno13_2}
A. Russomanno, A. Silva, and G. E. Santoro, J. Stat. Mech. P09012
(2013).
%
\bibitem{lazarides14}
A. Lazarides, A. Das, R. Moessner, Phys. Rev. E {\bf 90}, 012110 (2014).
%
\bibitem{citro15}
R. Citro, E. G. Dalla Torre, L. D'Alessio, A. Polkovnikov, M. Babadi, T. Oka, and E. Demler, arXiv:1501.05660v1 [quant-ph] (2015).
%
\bibitem{coddington1955}
E. A. Coddington and N. Levinson, {\em Theory of ordinary differential equations} (McGraw-Hill Book Company, Inc., New-York Toronto London, 1955).
%
\bibitem{moreno14}
M. P. Moreno and S. S. Vianna, Opt. Comm. {\bf 313}, 113 (2014).
%
\bibitem{zhu2000}
M. Zhu and L. S. Cutler, Proceedings of the 32nd Precise Time and Time Interval (PTTI) Meeting, pp. 311-323 (2000).
%



\end{thebibliography}
\end{document}